\documentclass[twoside]{dis07}
\usepackage[latin1]{inputenc}
\usepackage{graphics}
\usepackage{graphicx}
\usepackage{epsfig}
\usepackage{epstopdf}
\usepackage{wrapfig,rotating}
\usepackage{amssymb,amsmath,array}

\begin{document}
\title{Extracting the Gluon Piece of the Spin Puzzle\\ New Inclusive Jet Results from STAR}

\author{Renee Fatemi for the STAR Collaboration
\vspace{.3cm}\\
Massachusetts Institute of Technology\\
Laboratory of Nuclear Science \\
77 Massachusetts Ave. Cambridge, MA 02139 USA
}

\maketitle

\begin{abstract}

This contribution presents the most recent  mid-rapidity inclusive jet results from 3 $pb^{-1}$ of data collected from longitudinally polarized proton collisions at $\sqrt{s}=200$ GeV during the 2005 RHIC run. The inclusive jet asymmetry, $A_{LL}$,  with it's increased transverse momentum range and precision, provides strong constraints on the gluon helicity distribution when compared with existing next-to-leading order perturbative QCD evaluations.

\end{abstract}

\vspace{0.7cm}

\noindent Measurements of the partonic helicity distribution functions in the proton are essential for a complete understanding of the long range, non-perturbative properties of Quantum Chromodynamics (QCD).  Three decades of polarized lepton-nucleon deep-inelastic-scattering (DIS) experiments \cite{DIS} have shown that the probability for a quark spin to be aligned with the spin of the parent proton is $\sim{25}\%$.  Conservation of angular momentum requires the quark ($\Delta{Q}$) and gluon ($\Delta{G}$) total spin and orbital angular momentum ($L_Q$ + $L_G$) within the proton to sum to $\frac{\hbar}{2}$, motivating investigations into the size of the remaining components of the sum rule. Traditional fixed-target DIS experiments couple to the gluon distributions only at next-to-leading order (NLO), providing limited constraints on $\Delta{G}$\cite{SMCDIS}\cite{E155}. As a result,  several programs designed to directly access $\Delta{G}$ have been established and have produced initial results \cite{HERMES1}\cite{SMC}\cite{PHENIX1}\cite{COMPASS}\cite{STAR}.  Measurements of higher statistical precision and broader $Q^2$ reach continue to be recorded and released \cite{PHENIX2}.  During the 2005 RHIC run STAR sampled 3 $pb^{-1}$ of proton collisions with an average longitudinal beam polarization of 50$\%$ and $\sqrt{s}=200$ GeV.  As a result, the inclusive jet asymmetry measurement, spanning a transverse momentum ($p_T$) range of $5-32$ GeV, represents the most precise measurement over the largest $p_T$ range to date.

The Solenoidal Tracker at RHIC (STAR) Collaboration utilizes the polarized proton beam provided by the Relativistic Heavy Ion Collider (RHIC) to study final state interactions resulting from quark-quark (qq), quark-gluon (qg) and gluon-gluon (gg) scattering.  STAR's  large acceptance facilitates jet reconstruction, allowing for a nearly fragmentation independent reconstruction of the partonic characteristics inside the proton.  The inclusive jet double spin asymmetry $A_{LL}$,

\begin{equation}
A_{LL}=\frac{1}{P_YP_B}\frac{N^{++}-N^{+-}}{N^{++}+N^{+-}}=\frac{\sum_{AB\rightarrow{CX}}\Delta{f_A}\Delta{f_B}\Delta\sigma_{AB\rightarrow{CX}}}{\sum_{AB\rightarrow{CX}}{f_{A}f_{B}\sigma_{AB\rightarrow{CX}}}}
\end{equation}

\noindent is constructed from a ratio of the helicity aligned ($N^{++}$)  and anti-aligned ($N^{+-}$) luminosity normalized jet yields and is related at leading order to the  product of the initial polarized ($\Delta{f}$) and unpolarized ($f$) parton distribution functions and partonic polarized ($\Delta\sigma_{AB\rightarrow{CX}}$) and unpolarized ($\sigma_{AB\rightarrow{CX}}$) cross-sections.  The agreement between data and NLO perturbative QCD (pQCD) predictions for the inclusive jet cross-section\cite{STAR} motivate the ultimate extraction of the gluon helicity distribution from these asymmetry measurements. 

The STAR detector systems  \cite{STARDET} relevant for jet reconstruction are the Time Projection Chamber  (TPC) and the Barrel (BEMC) and Endcap (EEMC) Electromagnetic Calorimeters.  The TPC provides momentum information on charged particles scattered in the pseudorapidity region $|\eta|<1.4$.   The BEMC and EEMC measure the neutral energy deposited per event in the range spanning $-1<\eta<2$. The minimum bias trigger (MINB) was defined by a coincidence signal between East and West Beam Beam Counters (BBC) \cite{BBC}, which are  segmented scintillator detectors located on either side of the interaction region at $3.3 <|\eta|<5.0$. Enhancement of jet reconstruction at  high transverse momentum ($p_T$)  was achieved by requiring a MINB plus High Tower (HT) or Jet Patch (JP) trigger to be fullfilled.  The HT trigger required a single BEMC tower ($\Delta{\phi}=\Delta{\eta}=0.05$) to exceed a low/high threshold of 2.6/3.5 GeV. The JP trigger, included for the first time in this analysis, required that a cluster of towers ($\Delta{\phi}=\Delta{\eta}=1.0$) exceed the low/high threshold of 4.5/6.5 GeV. Finally the BBC also serves as the STAR luminosity monitor, providing the spin dependent luminosity normalization for the asymmetry analysis. 

\begin{figure}[h]
\begin{center}
\includegraphics[height=6cm,width=9cm]{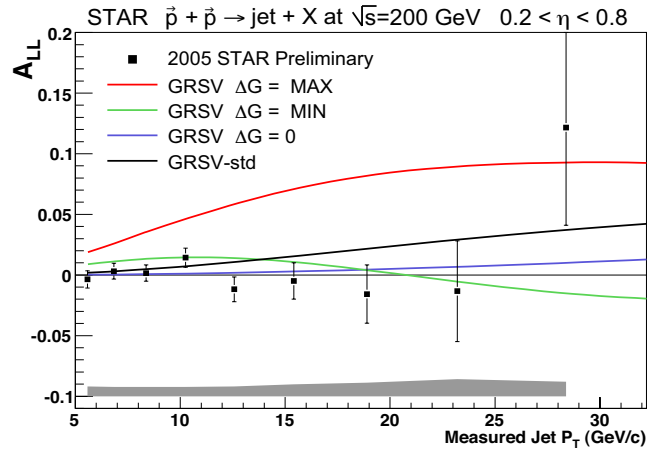}
\caption{\label{ALL} The 2005 STAR mid-rapidity inclusive jet asymmetry calculated from from 3 $pb^{-1}$ of sampled data. The error bars are statistical only with systematic errors represented by the gray shaded band as the bottom. The curves are NLO pQCD calculations\cite{JETGRSV}  depicting the inclusive jet asymmetry for various input $\Delta{G}$ distributions. }
\end{center}
\end{figure}

The  STAR jetfinder is based on the mid-point cone algorithm \cite{ALGO}, which clusters TPC tracks and BEMC tower energies greater than 0.2 GeV into jets of radius $R=\sqrt{\Delta\phi^2+\Delta\eta^2}=0.4$,  requiring a split/merge fraction of 0.5 and seed energy of 0.5 GeV. Jet reconstruction is limited to the region of the BEMC which was fully implemented and incorporated into the trigger during the 2005 run, resulting in the requirement that the jet axis lies at least 0.2 units in pseudorapidity from the detector edge $( 0.2 < \eta_{jet} < 0.8)$.  Beam background, not associated with the hard scattering, results in an excess of neutral energy from the calorimeters to be clustered into the reconstructed jet. The sub-sample of jets dominated by this beam background are removed by requiring the neutral energy fraction of the jet  be less than 0.8.  Additionally, the jets were required to originate from a vertex of $\sim\pm 60$ cm along the beamline.
   
Figure \ref{ALL} shows the 2005 STAR inclusive jet $A_{LL}$ as a function of the measured jet $p_T$. Although the HT and JP trigger rates were matched in bandwidth, the high threshold JP trigger jet reconstruction efficiency was on average $50\%$ greater than the HT, resulting in half of the final jet sample originating from the higher threshold JP trigger. In order to maintain a uniform trigger bias, each jet was required offline to contain a trigger tower (HT) or patch (JP). The grey shaded band represents the systematic error, excluding the 25$\%$ uncertainty on the beam polarization values. The leading contributions to the systematic error result from an estimate of the bias introduced to the asymmetries from the trigger and jet reconstruction requirements and a conservative upper limit on possible false asymmetries in the measurement. Additional, less significant, systematic contributions arise from non-longitudinal beam components and the effect of beam background on the asymmetry and relative luminosity measurements.  No significant bunch or fill dependent systematics were observed.  The 2005 jet asymmetries are in good agreement with previous STAR inclusive jet measurements \cite{STAR} in the region of kinematic overlap.

\begin{figure}[ht]
\begin{center}
\includegraphics[height=7cm,width=7.cm]{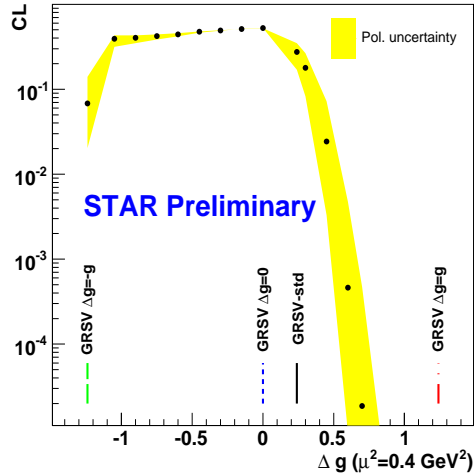} 
\caption{\label{CL}Confidence level comparisons between the 2005 STAR inclusive jet asymmetry measurement and NLO pQCD calculations for various gluon distributions within the GRSV framework\cite{GRSV}\cite{JETGRSV}. The yellow shaded band represent the systematic error due to the beam polarization uncertainty.}
\end{center}
\end{figure}

The curves in Fig.\ref{ALL} are NLO pQCD calculations of inclusive jet asymmetries\cite{JETGRSV} for several variations of $\Delta{G}$ at the input scale. The black curve (standard) incorporates the $\Delta{G}$ which gives a best fit to the DIS data within the GRSV formulation\cite{GRSV}.  The red (green) curves indicate scenarios where the gluons are completely (anti-)aligned  with the proton spin. The blue curve shows $\Delta{G}=0$ and therefore all non-zero values derive from the quark spins only. The jet asymmetries are clearly not consistent with the maximal polarization case, but are also not yet precise enough to distinguish between the minimal, standard and zero scenarios.  The precision of the measurement does allow, within the GRSV framework, for the exclusion of several $\Delta{G}$ values between the standard and maximal case. Figure \ref{CL} quantifies this comparison by showing the confidence level at which this measurement rules out various input gluon distributions.  The 2005 results rule out $\Delta{G} >$ 0.5 at $\sim95\%$ level. In comparison, the $\chi^2=\pm1$ range for  the GRSV standard gluon fit spanned $\Delta{G}= -0.45$ to $0.7$.   Systematic errors and their correlations across $p_T$ bins are accounted for in the confidence level calculations, with the yellow shaded band representing the possible systematic shift due to the beam polarization uncertainty.  The $p_T$ reach of this measurement translates into a sampling of $\sim{50}\%$ of the total integral  ($x_{Bjorken} = 0.03-0.3$) of the gluon helicity distribution. 

The STAR 2005 inclusive jet results provide a significant contribution to the global understanding of $\Delta{G}$. Although the measurements presented here clearly place stronger constraints on the gluon distribution than the fixed target DIS data, only their inclusion within a global analysis of world data will allow for a definitive extraction of $\Delta{G}$.     
     
\section*{Acknowledgements} The authors would like to thank the organizers of the conference for the opportunity to present these results. We would also like to thank  for W. Vogelsang and M. Stratmann for providing additional curves of the inclusive jet asymmetries with varying input gluon polarization values.      

\begin{footnotesize}

\end{footnotesize}

\end{document}